\documentclass[twocolumn]{IEEEtran}

\usepackage{amsmath}
\usepackage{multicol}
\normalsize

\newcommand{\req}{\text{req}}
\newcommand{\mb}{\mathbf}
\newcommand{\mc}{\mathcal}
\newcommand{\F}{\mathcal{F}_{\text{req},t}}
\newcommand{\B}{\mathcal{B}}

\newcommand{\E}{\mathbf{E}}

\newcommand{\SINR}{\text{SINR}}
\newcommand{\FH}{R_{f,t}^{\text{FH}}}

\usepackage{amsfonts,upgreek,pifont,amssymb,enumerate,color,algorithm,theorem}

\usepackage{algpseudocode}
\usepackage{cite,graphicx,subfigure,epstopdf}

\usepackage{mathtools}

\newenvironment{proof}
    {
      \\\emph{Proof.}
    }
    {

    }

\def\BibTeX{{\rm B\kern-.05em{\sc i\kern-.025em b}\kern-.08em
    T\kern-.1667em\lower.7ex\hbox{E}\kern-.125emX}}

\begin{document}
\bstctlcite{IEEEexample:BSTcontrol}
\title{Joint Long-Term Cache Updating and Short-Term Content Delivery in  Cloud-Based Small Cell Networks}

\author{
Xiongwei~Wu, {\it Student Member, IEEE}, Qiang Li, {\it Member, IEEE}, Xiuhua Li, {\it Member, IEEE}, \\Victor C. M. Leung, {\it Fellow, IEEE}, P. C. Ching, {\it Fellow, IEEE}
\thanks{
Part of this work was presented at the IEEE International Conference on Communications (ICC), 20-24 May 2019, Shanghai, China.}%
\thanks{X. Wu and P. C. Ching are with the Department of Electronic Engineering, Faculty of Engineering, The Chinese University of Hong Kong, Shatin, Hong Kong SAR of China  (e-mail: xwwu@ee.cuhk.edu.hk; pcching@ee.cuhk.edu.hk).}
\thanks{Q. Li is with the School of Information and Communication Engineering, University of Electronic Science and Technology of China, Chengdu 611731, China, and also with the Peng Cheng Laboratory, Shenzhen 518052, China (e-mail: lq@uestc.edu.cn).}
\thanks{X. Li is with the School of Big Data and Software Engineering, Chongqing University, Chongqing 401331, China (e-mail: lixiuhua1988@gmail.com).}
\thanks{V. C. M. Leung is with the College of Computer Science and Software Engineering, Shenzhen University, Shenzhen 518060, China, and also with the Department of Electrical and Computer Engineering, The University of British Columbia, Vancouver, BC, V6T 1Z4 Canada (email: vleung@ieee.org).}
}
\maketitle

\begin{abstract}
Explosive growth of mobile data demand may impose a heavy traffic burden on fronthaul links of cloud-based small cell networks (C-SCNs), which deteriorates users' quality of service (QoS) and requires substantial power consumption. This paper proposes an efficient maximum distance separable (MDS) coded caching framework for a cache-enabled C-SCNs, aiming at reducing long-term power consumption while satisfying users' QoS requirements in short-term transmissions. To achieve this goal, the cache resource in small-cell base stations (SBSs) needs to be reasonably updated by taking into account users' content preferences, SBS collaboration, and characteristics of wireless links. Specifically, without assuming any prior knowledge of content popularity, we formulate a mixed timescale problem to jointly optimize cache updating, multicast beamformers in fronthaul and edge links, and SBS clustering. Nevertheless, this problem is anti-causal because an optimal cache updating policy depends on future content requests and channel state information. To handle it, by properly leveraging historical observations, we propose a two-stage updating scheme by using Frobenius-Norm penalty and inexact block coordinate descent method. Furthermore, we derive a learning-based design, which can obtain effective trade-off between accuracy and computational complexity. Simulation results demonstrate the effectiveness of the proposed two-stage framework. 
\end{abstract}

\IEEEpeerreviewmaketitle

\section{Introduction}
Recently, cloud-based small cell networks (C-SCNs) have been widely regarded as a promising mechanism for fifth generation (5G) wireless networks \cite{bhushan2014network}. Under the coordination of the cloud processor (CP), C-SCNs enables seamless collaboration among small base stations (SBSs) and centralized optimization for resource allocation and signal processing \cite{zhang2016fronthauling,hardjawana2016parallel}. However, the unprecedented growth of mobile data demand, caused by satisfying various service requests from mobile users in recent years, can impose a heavy traffic burden on fronthaul links of the C-SCNs and degrade the quality of service (QoS) for users. Moreover, such a traffic explosion may also require substantial fronthaul energy consumption with the progressively dense deployment of SBSs \cite{zhang2016fronthauling}. This issue will play an increasingly important role in green communication \cite{chih2014toward,hardjawana2016parallel} since energy consumption of wireless networks not only brings economic costs to mobile network operators (MNOs), but also has negative ecological and health impacts on human beings. To cope with these challenges, caching at wireless edge nodes in the C-SCNs is considered to constitute an effective approach towards 5G wireless networks \cite{bastug2014living}.


In a cache-enabled C-SCNs, each SBS is equipped with a local cache that is capable of storing a limited number of popular contents \cite{tao2016content}. Thus, when these cached contents are requested by users, they can be directly accessed in local SBSs without fronthaul transmissions. Massive duplicated transmissions via fronthaul and backhaul links can be avoided, which leads to a tremendous reduction in traffic loads \cite{li2018resource}. In addition, pushing contents closer to mobile users and frequently reusing cached contents can assist to significantly decrease access latency of mobile users, transmit power of SBSs, and energy costs of fronthaul \cite{golrezaei2012femtocaching}. Therefore, edge caching can fundamentally improve the network performance of C-SCNs \cite{li2018resource}.

Edge caching has been investigated in a vast amount of literature. In cache-enabled networks, one critical issue is {\bf how to design long-term caching strategies in terms of improving network performance}, since users' content preferences usually evolve fairly slowly, i.e., on the order of hours or even days. In most extant literature, the caching schemes were designed in an offline manner with prior knowledge of content popularity distribution (e.g., the well-known Zipf distribution \cite{liao2017coding,li2018hierarchical}).
The authors in \cite{golrezaei2012femtocaching} first came up with the idea of Femtocaching to minimize download delay. The studies in \cite{hsu2016resource,li2015delay,bacstuug2016delay} examined effective caching strategies in order to decrease access latency when satisfying mobile users' content requests.
However, all of these studies mainly focused on optimizing averaged network performance metrics and caching strategies, ignoring the last mile of physical transmissions (i.e., wireless transmissions of content delivery from BSs to mobile users).
Another critical issue for cache-enabled wireless networks is {\bf how to schedule content delivery policy so as to satisfy users' requests over a much shorter timescale}, i.e., considering signal processing and resource allocation. To take advantage of cooperative beamforming introduced by edge caching, the studies in \cite{wu2019latency,wu2019JointFronthaul,tao2016content,peng2017layered,Jaber2018ICC} investigated content delivery policy by optimizing beamformers and SBS clustering for content downloads. However, these preliminary works generally studied cooperative beamforming under fixed or heuristic caching schemes. Hence, the benefits of cooperative beamforming in cached-aided wireless networks have not yet been fully elucidated.

Apparently, the aforementioned two issues in cache-enabled wireless networks are highly coupled. Specifically, in order to allow cached contents to be repeatedly used, the contents available in the storage-limited SBSs should be updated by taking physical-layer transmissions into consideration, e.g., cooperative pattern of SBSs and wireless links. Meanwhile, the performance of cooperative beamforming during content delivery is also highly affected by dynamic cache resources in SBSs \cite{wu2019JointFronthaul}. Herein, in this paper, to fully exploit the benefits of edge caching and cooperative beamforming, we investigate effective content caching design by jointly considering long-term cache updating and short-term content delivery. The goal of the proposed content caching design is to minimize long-term power consumption while satisfying mobile users' content requests with QoS guarantees. To achieve this goal, we are motivated to formulate a mixed timescale optimization problem and propose an efficient content caching framework from the perspective of engineering implementation.

There are some key features that distinguish the proposed caching framework from extant studies. 
{In this study, cache resource at SBSs is envisioned to be updated by learning and tracking users' content preferences, collaborative patterns among SBSs, and characteristics of wireless links. Instead of assuming any prior distribution of content popularity \cite{hsu2016resource,li2015delay,bacstuug2016delay,bioglio2015optimizing,liao2017coding,li2018hierarchical}, the proposed solution is as a result of {\it historical observations} of users' requests and channel state information (CSI). We should mention that there are some other studies assuming no prior distribution of content popularity, e.g., \cite{abedini2014content,liu2019mixed}. Nevertheless, our proposed caching framework still has some distinguishing features. 
Specifically, the work in \cite{abedini2014content} did not address the issue of beamforming and SBS clustering for content delivery in physical-layer transmissions. 
Moreover, the research focus in \cite{liu2019mixed} was limited into two transmission modes only, i.e. each user is served by either one SBS or all SBSs. 
In contrast, the proposed framework is capable of pushing popular contents into different SBS clusters so as to enhance cooperative beamforming gain.  
Another key feature is exploiting fronthaul multicast enabled by maximum distance separable (MDS) codes to enhance SBS collaboration and reduce power consumption. Specifically,  uncached MDS coded contents can be delivered to the associated SBS cluster via fronthaul using the same communication resource.
The studies in \cite{tao2016content,peng2017layered} adopted only uncoded caching and fronthaul unicast to minimize the power consumption, which may suffer from high fronthaul cost. The study in \cite{golrezaei2012femtocaching} applied MDS codes to reduce download delay without investigating a joint design of cooperative beamforming and fronthaul multicast.} Meanwhile, the utilization of coded caching motivates us to design multicast beamformers in fronthaul and edge links simultaneously, which differs from most existing works that consider multicast beamforming only in edge transmissions.
Finally, under MDS caching, the condition for avoiding service interruption is also incorporated into our problem, which can further enhance the QoS of users. 
However, these features also lead to a very challenging problem in jointly designing long-term cache updating and short-term content delivery.

With regard to joint design of long-term cache updating and short-term transmission policy, only a few preliminary designs exist. 
For instance, cache updating schemes 
were studied in some learning-based research \cite{song2017learning,chattopadhyay2018gibbsian,azimi2018online,sadeghi2018optimal}. However, the statistical models used in these studies, e.g., the Markov decision process (MDP) model, generally make it difficult to include signal processing for content delivery, i.e., cooperative beamformer design, (signal-to-interference-plus-noise ratio) SINR constraints, etc. 
From the perspective of secure transmission, the work in \cite{xiang2018cache} investigated beamformer design for edge links while adopting fronthaul unicast transmissions; the benefits of using coded caching and fronthaul multicast were not investigated. 
Moreover, our preliminary study in \cite{Wu2019ICC} also adopted an uncoded caching scenario and studied content delivery via fronthaul unicast transmissions. Thus, the issues of employing coded caching, multicast beamforming in both fronthaul and edge links, and efficient framework design still need to be further investigated.

The main contributions of this paper are summarized as follows:
\begin{itemize}
	\item We propose an efficient content caching framework towards future green  wireless networks by integrating MDS coded caching, multicast beamforming in fronthaul and edge links, and SBS collaboration, with the goal of reducing long-term power consumption while satisfying users' content requests with guaranteed QoS in short-term transmissions. In particular, without assuming any priori distribution, the cached contents available in SBSs are enabled to be reasonably updated by considering users' requests and potential collaborative patterns of SBSs as well as wireless links. Moreover, under MDS coded caching, we incorporate fronthaul rate and SINR constraints to avoid interrupted downloads and enhance users' QoS.
	\item By properly making use of historical observations, we develop a two-stage updating scheme to handle the formulated mixed timescale optimization problem with regard to short-term content delivery design and long-term cache updating design from the perspective of engineering implementation. The former subproblem constitutes a mixed integer nonlinear program (MINLP) and is handled by using a Frobenius-Norm penalty approach, while the latter one is solved by a distributed algorithm. Furthermore, we propose another efficient algorithm to reduce computational complexity by learning local content preferences of the users served by each SBS.
	\item Numerical results are provided to demonstrate the effectiveness of the proposed framework, especially concerning the importance of how joint design of cache updating and cooperative beamforming reduces energy cost and how cooperative beamforming augments the efficiency of content delivery under the fronthaul multicast enabled by MDS codes.
\end{itemize}
The remainder of this paper is organized as follows. Sec. II introduces the system model. Sec. III presents problem formulation and decomposition. Sec. IV elaborates on the proposed algorithm for short-term content delivery design and Sec. V introduces practical implementation of the two-stage content-caching framework. Sec. VI presents the performance evaluation for the proposed scheme. Finally, Sec. VII concludes the paper.

\section{System Model}
\begin{figure}[h]
  \centering
  \includegraphics[scale=0.6]{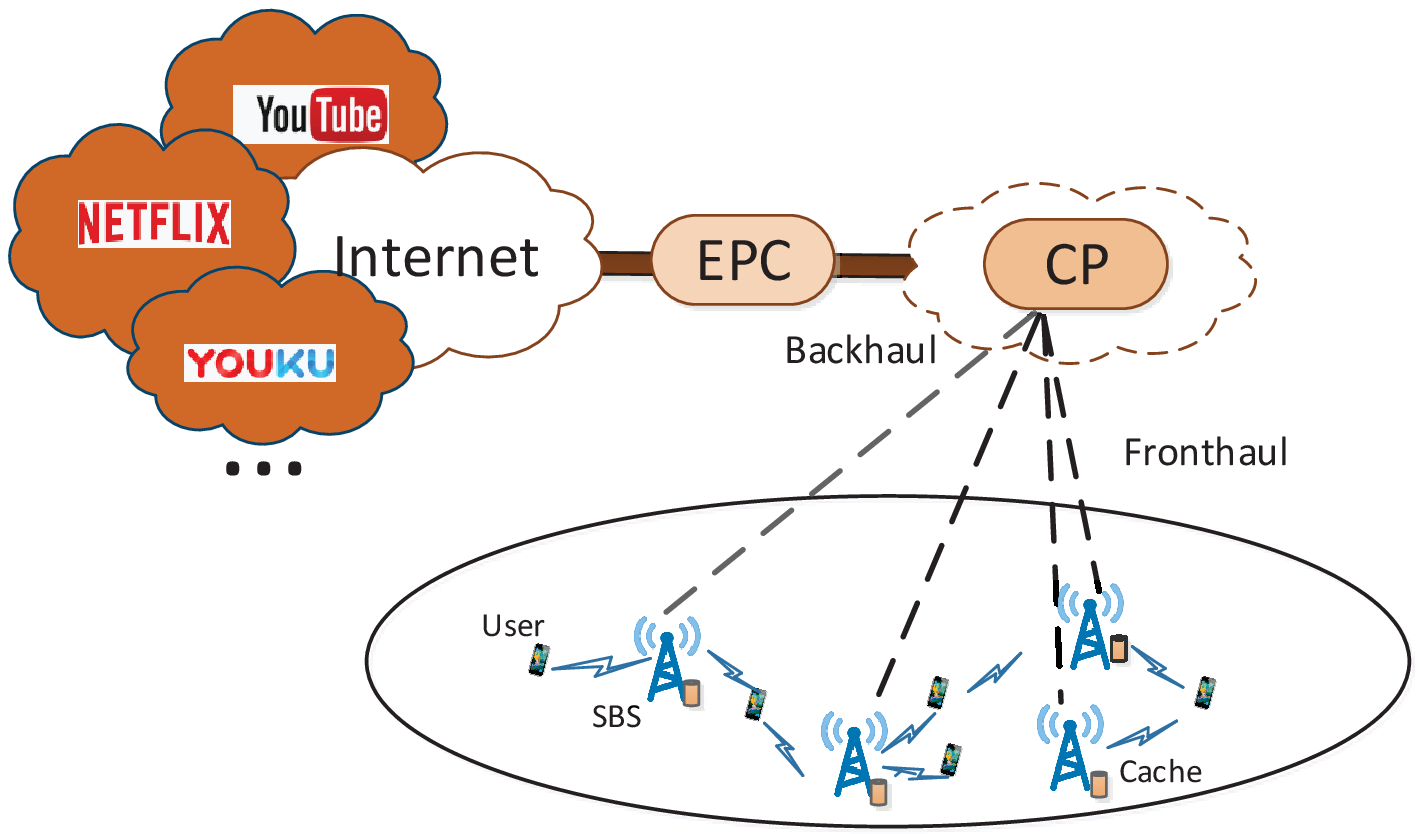}
  \caption{An illustration of cache-enabled C-SCNs. }
  \label{system}
\end{figure}

As depicted in Fig. \ref{system}, in the considered cache-enabled C-SCNs, a total of $B$ densely deployed SBSs are connected to the CP through wireless fronthaul links. The CP has $N$ antennas, while each SBS $b$ is equipped with $M$ antennas and a cache unit with the storage size of $S_b$ bits.
There are $K$ randomly geographically distributed single-antenna users in the C-SCNs. The active users can be cooperatively served by a cluster of SBSs via wireless cellular links (also referred to as edge links) under the coordination of the CP.
Moreover, by using backhaul links, the CP is connected to the Enhanced Packet Core (EPC) and the EPC is further connected to the Internet. Thus, the CP can access the considered whole library of $F$ popular contents from various content providers (e.g., YouTube, Netflix, etc.). In addition, each content $f$ has a size of $s_f$ bits. Let sets ${\cal{B}} = \{1,\dots, B\}$, $\mathcal{F} = \{1,\dots, F\}$, and $\mc K = \{1,\dots, K\}$ be the indices of SBSs, contents in the library and users, respectively. The key notations of this paper are listed in Table \ref{table:1}.

\begin{table}[h!]
\centering
\caption{Key notations}
\begin{tabular}{|c|c|}
 \hline
  $\mc B_{f,t}$ & SBS cluster for cooperatively delivering content $f$ at frame $t$\\
 \hline
 $\F$ & set of requested contents at frame $t$\\
 \hline
 {$\mc G_{f,t}, \mc K$} & {set of users requesting content $f$ at frame $t$, set of all users}\\
 \hline
 ${\mb L}$ & {cache allocation matrix}\\
 \hline
 ${\mb E_t}$ & {SBS clustering matrix at frame $t$}\\
 \hline
 ${\mc W_t, \mc V_t }$ & {aggregate beamformers for fronthaul and edge links at frame $t$}\\
 \hline
\end{tabular}
\label{table:1}
\end{table}

We consider heterogeneous users' content preferences \cite{liao2017coding}, i.e., different users may have distinct preferences towards these popular contents.
In most previous proactive caching studies, the caching strategies were investigated under the perfect knowledge of users' preference patterns and content popularity distributions, such as the well-known Zipf distribution \cite{liao2017coding}. Specifically, the frequency of content $f$ being selected by users in preference pattern $i$ is given by
    $p_{i,f} = c_i \zeta_{i,f}^{-\kappa_i}$,  
where $\kappa_i$ denotes the skewness factor; $\zeta_{i,f}$ denotes the rank order of content $f$ in preference pattern $i$; and $c_i$ is a factor for normalization. It is worth noting that these parameters may not be available as prior information in practical engineering implementations.
Hence, different from conventional proactive caching designs, we do not pre-specify any knowledge of content popularity in this work. 

To proceed, the CP needs to make transmission policy when users' requests arrive. Specifically, we consider block fading scenarios, and the process of content delivery is assumed to operate in a transmission frame; 
moreover, the wireless channel is assumed to have relatively long coherence time (e.g., for very low mobility scenario) so that during the transmission the channel is static. 
Accordingly, transmission beamformers and SBS collaboration need to be optimized based on the current contents cached in SBSs, instantaneous CSI, etc., which constitutes a \textbf{short-term} process. Another task for the CP is to coordinate all SBSs and update the cached contents in order to achieve high cooperative gains. In realistic scenarios, the content popularity distribution generally varies slowly (e.g., on the order of hours \cite{li2018hierarchical}), and the update of cached contents may not be very frequent so as to reduce overload cost and complexity \cite{wang2016mobility}. Thus, cache updating is generally a \textbf{long-term} process. Specifically, at each cache-updating frame $t_0$, the cached contents should be reasonably updated so that they can be repeatedly accessed in a long timescale $\mc T =\{t_0 + 1, \cdots, t_0 + T\}$. This process can lead to a significant reduction in fronthaul energy costs. 

{Moreover, in the C-SCNs, the CP can effectively update the cached contents by learning and tracking the underlying collaborative pattern of SBSs, users' content preferences, and the quality of wireless links. The detailed models for MDS coded caching, edge and fronthaul delivery, and power consumption are introduced as follows.}

\subsection {MDS Coded Caching Model}
To provide abundant opportunities for SBS cooperation and fronthaul multicast, we adopt MDS codes to encode contents \cite{liao2017coding}. Specifically, each content is divided into multiple segments with an equal size of $s_0$ bits.
With the technique of MDS codes, each information segment can be encoded into a sufficiently long sequence of parity bits so that any $s_0$ parity bits are sufficient to recover the original information segment \cite{liu2014cache,liu2015exploiting}.
Herein, we denote $l_{f,b} \in [0,1]$ as the cache allocation variable for the fraction of content $f$ to be cached in SBS $b$. Similar to \cite{liu2015exploiting}, for any content $f$, SBS $b$ is assumed to randomly store $l_{f,b}s_f$ parity bits of these MDS coded segments. 
Notably, when any content $f$ is not fully cached in SBS cluster $\mc B_0$, i.e., $l_{f,b} < 1, \forall b \in \mc B_0$, these SBSs need to fetch the missing parity bits from the CP. On the basis of MDS decoding, the CP can always transmit $\max_{b \in \mc B_0} (1 - l_{f,b})s_f$ uncached parity bits to these SBSs $\mc B_0$ via fronthaul multicast so that content $f$ can be reconstructed in each SBS \cite{liao2017coding}. For notational convenience, we denote $\mb L= [l_{f,b}] \in \mathbb{R}^{F\times B}$ as the cache allocation matrix for the considered period $\mc T$.

\subsection {Edge Transmission Model}
At the beginning of each frame $t$, the requests from all active users, denoted by $\pi_t = \{(k,f)|k \in \mc K_t, f \in \F\}$, are revealed,
where $\mc K_t \subseteq \mc K$ and $\mc F_{\req,t} \subseteq \mc F$ are the sets of active users and requested contents at frame $t$, respectively. Consider that each active user only requests one content in each frame. Subsequently, the users requesting the same content (i.e., content $f$) are formed as a multicast group (denoted by $\mc G_{f,t}$), and thus served by a cluster of SBSs (denoted by $\mc B_{f,t}$) through multicast beamforming.
We define the SBS clustering matrix at frame $t$ as
\begin{align}
   \E_t = [ e_{f,b,t}] \in \{ 0,1 \}^{F_{\req,t} \times B }, \label{p1e}
\end{align}
where ${F}_{\req,t}$ denotes the cardinality of $\mc F_{\req,t}$; and the element $e_{f,b,t} = 1 $ implies that SBS $b$ is scheduled for transmitting content $f$ to the users in multicast group $\mc G_{f,t}$ at frame $t$; otherwise, $e_{f,b,t} = 0$.
Therefore,
the transmitting signal from SBS $b$ at frame $t$ is
    \begin{align}
      {{\mathbf{x}}_{b,t}} = \textstyle\sum_{f \in \F} {{{\mathbf{v}}_{f,b,t}}{x_{f,t}}}, \forall b\in \mc B, 
    \end{align}
where $\mathbf{v}_{f,b,t} \in \mathbb{C}^M$ denotes the transmit beamformer from SBS $b$ to precode information symbol $x_{f,t} \in \mathbb{C}$ for multicast group $\mc G_{f,t}$ with $\mathbb{E}[|x_{f,t}|^2] = 1$. We further define $\mc V_t =\{\mb v_{f,b,t}, \forall f \in \F, \forall b \in \mc B\}$.
The total transmit power of SBS $b$ is limited by
\begin{align}
  \textstyle\sum_{f \in \F} \mb \|\mb v_{f,b,t}\|^2 \leq P_b, \forall b \in \B, \label{p1d}
\end{align}
where $P_b$ is the maximum transmit power of SBS $b$.
Particularly, if SBS $b$ is not selected to serve the multicast group $\mc G_{f,t}$ at frame $t$, then the associated transmit beamformer $\mathbf{v}_{f,b,t}$ equals $\mathbf{0}$, i.e.:
\begin{align}
(1 - e_{f,b,t})\mathbf{v}_{f,b,t} = \mathbf{0}, ~\label{p1c}  \forall f\in \F, \forall b\in \mc B.
\end{align}
We consider the frequency-flat fading wireless channels, and the receiving signal at user $k$ that requests content $f$ at frame $t$ can be expressed as
  \begin{align}
     y_{k,t} = \underbrace{\mb h_{k,t}^H \mb v_{f,t} x_{f,t}}_{\text{desired signal}} + \underbrace{\sum_{f' \in \mc F _{\req,t} \backslash \{f\}} \mb h_{k,t}^H \mb v_{f',t} x_{f',t}}_{\text{inter-group~interference}} + n_{k,t}, \forall k\in \mc G_{f,t},
  \end{align}
where the aggregate channel matrix ${{\mathbf{h}}_{k,t}} = \left[ {{\mathbf{h}}_{k,1,t}^H,{\mathbf{h}}_{k,2,t}^H, \cdots,{\mathbf{h}}_{k,B,t}^H}\right]^H$ denotes the CSI from all SBSs to user $k$, and the channel vector ${{{\mathbf{h}}_{k,b,t}}} \in \mathbb{C}^{M}$ denotes the CSI between SBS $b$ and user $k$; the aggregate beamformer ${{\mathbf{v}}_{f,t}} = \left[ {{\mathbf{v}}_{f,1,t}^H,{\mathbf{v}}_{f,2,t}^H,\cdots,{\mathbf{v}}_{f,B,t}^H} \right]^H$ denotes the beamformers from all SBSs to precode signal $x_{f,t}$; and ${n}_{k,t}$ denotes the additive complex Gaussian noise, i.e.,  ${n_{k,t}} \sim \mathcal{CN}({0},\sigma_{k}^2)$, at frame $t$. Therefore, the received signal-to-inference-plus-noise ratio (SINR) for user $k$ requesting content $f$ at frame $t$ can be calculated as:
\begin{align}
   \SINR_{k,t} = {|\mb h_{k,t}^H \mb v_{f, t}|^2}/g_{k,t}({\mc V_t}),
\end{align}
where $g_{k,t}({\mc V_t}) \triangleq \sum_{f' \in \F \backslash \{f\}} |\mb h_{k,t}^H\mb v_{f',t} |^2 + \sigma_{k}^2$.

To guarantee user QoS, the minimum SINR $\gamma_{f}$ at each user $k$ in multicast group $\mc G_{f,t}$ needs to be satisfied as
\begin{align}
   \text{SINR}_{k,t} \geq \gamma_{f}, \forall k \in \mc G_{f,t}, \forall f \in \F. \label{p1b}
\end{align}
Thus, a feasible transmission rate $R_f$ for delivering content $f$ can be given by $R_f = B_1\log_2 (1 + \gamma_f)$, where $B_1$ (Hz) denotes the bandwidth of edge links \cite{tao2016content}.

\subsection {Fronthaul Transmission Model}
When the requested contents are not entirely stored in local SBSs, the uncached portion of contents (parity bits) needs to be fetched from the CP via fronthaul.
To capture the benefits of MDS codes, we also adopt the multicast beamforming technique in the fronthaul transmission. To avoid interference between fronthaul and edge links, the radio spectrum of the fronthaul links is orthogonal to that of the edge links. Let $\underline x_{f,t}$ be the information symbol that encodes content $f$ in the CP at frame $t$, and $\mathbb{E} [|\underline x_{f,t}|^2] = 1$. Hence, the receiving signal\footnote{For simplicity, we consider a single data stream for each SBS in fronthaul transmissions. The idea of the proposed design also works by adopting multiple data streams.} at SBS $b$ is
\begin{align}
	\underline {\mb y}_{b,t} = \mb H_{b,t}^H \mb w_{f,t}\underline x_{f,t} + \underline{\mb z}_{b,t},
\end{align}
where $\mb w_{f,t} \in \mathbb C^{N} $ denotes the multicast beamformer that precodes information symbol $\underline x_{f,t}$; $\underline{\mb z}_{b,t}$ denotes the additive Gaussian noise with distribution $\mc{CN} (\mb 0, z_b^2 \mb I)$; and the channel matrix $\mb H_{b,t} \in \mathbb{C}^{N \times M}$ denotes the CSI between the CP and SBS $b$ at frame $t$. We also denote $\mc W_t = \{\mb w_{f,t}, \forall f \in \F\}$.
For simplicity, different fronthaul frequencies are allocated to different SBS clusters. Hence, by applying the technique of maximum ratio combining (MRC) \cite{hu2017joint}, the fronthaul multicast data rate for SBS cluster to serve group $\mc G_{f,t }$ is given by
\begin{align}
  \!R_{f,t}^{\text{FH}} = \min_{b \in \mc B_{f,t}} B_2 \log_2 (1 + \|\mb H_{b,t}^H \mb w_{f,t}\|^2/z_b^2), \forall f \in \F, \label{eq:mulrat}
\end{align}
where $\mc B_{f,t} = \{b| e_{f,b,t} = 1, \forall b \in \mc B\}$ denotes the SBSs that are selected to deliver content $f$ at frame $t$; and $B_2$ is the bandwidth for each SBS cluster, {and the total fronthaul bandwidth is assumed to be larger than $B_2K$.} It is worth noting that the multicast data rate is dominated by the channel capacity of the SBSs with the worst channel gain \cite{hu2017joint}. Hence, SBS clustering should depend on the channel gains of fronthaul and edge links simultaneously. If coordinating one SBS with a poor fronthaul link or edge link to deliver the content, it would unnecessarily consume more power to satisfy the requirements of user QoS.

Recall that \eqref{p1b} is provided to guarantee the throughputs of users. Nevertheless, this SINR constraint is still not sufficiently effective to ensure the requirements of user QoS in content-centric wireless networks.
For instance, when users request videos online, they will not expect to be interrupted or experience a long buffering time once they have started watching videos. 
To avoid these buffering issues, the SBSs should always store the next segment before it is delivered to users. 
{The study in \cite{vu2018latency} investigated the buffering time for content delivery in uncoded caching. Specifically, for uncoded caching scheme, the buffering time for delivering content $f$ at SBS $b$ is given by $t_f = [ {s_f/R_f} - {s_f(1-l_{f,b})}/{R_{f,t}^{\rm FH}}]^+$, where operator $ [\cdot]^+ = \max \{\cdot, 0\}$. Notably, if zero buffering time is experienced, i.e., $t_f = 0$, it gives rise to ${R_{f,t}^{\rm FH}} \geq (1-l_{f,b})R_f$ \cite{vu2018latency}. This event is considered as no {\it service interruption} during content delivery in this paper. Prior work \cite{peng2017layered} considered this fronthaul constraint for beamformer design.
For the MDS coded caching scheme, we present the following proposition to further enhance user QoS.
\vspace*{0.2cm}\\
{\it {\bf Proposition 1:}
Under MDS coded caching, it needs to satisfy
\begin{align}
  R_{f,t}^{\text{FH}} \geq \max_{b \in \B} ~ (1 - l_{f,b})e_{f,b,t}R_f,~ {\forall f\in   \F}, \label{eq:FH}
\end{align}
to avoid service interruption during content delivery. 
}
}
\vspace*{0.2cm}
\begin{proof}
See Appendix A.  \hfill $\blacksquare$
\end{proof}

By Proposition 1, we observe that when the requested content (e.g., content $f$) is fully placed in the selected SBSs, i.e., $l_{f,b} = 1$ and $e_{f,b,t} = 1$, its fronthaul transmissions are not needed, i.e.,  $R_{f,t}^{\text{FH}} = 0$ and $\|\mb w_{f,t}\|^2 = {0}$. Consider another special case in which $l_{f,b} = 0$ and $e_{f,b,t} = 1$. In this case, \eqref{eq:FH} can be simplified as $ R_{f,t}^{\text{FH}} \geq R_f$, which eventually leads to  maximum fronthaul power consumption. Intuitively, the fewer contents are cached in local SBSs, the higher fronthaul data rate is required to satisfy user QoS at a cost of more fronthaul power consumption. 
Moreover, compared with the fronthaul capacity constraint used in uncoded caching studies \cite{peng2017layered}, the derived condition in Proposition 1 under MDS coded caching is more challenging to address due to the involvement of fronthaul multicasting and max operator.

\subsection {Power Consumption Model}
The overall power consumption of a cached-enabled system generally arises from three parts, i.e., system maintenance, cache updating, and content delivery \cite{gabry2016energy,peng2017layered}. Specifically, power consumption for system maintenance is usually utilized to keep SBSs and the CP active, which is generally a fixed cost for a deployed network \cite{gabry2016energy}. This can normally be decreased by turning a few SBSs off, which is not considered here in order to simply the design. In practice, the cached contents usually remain fixed for a long term after they are stored in SBSs
while users' requests arrive much more frequently. Thus, the frequent reuse of cached contents can reduce considerable fronthaul power consumption, which leads the energy costs for cache updating to be negligible \cite{peng2017layered}. Therefore, we focus on power consumption for content delivery in this paper.

Power consumption during content delivery mainly comprises two parts: edge transmission power from SBSs to users; and fronthaul transmission power when the CP transmits the uncached MDS coded contents to SBSs via multicast transmissions.
Accordingly, the content delivery power consumption at frame $t$ is given by
   \begin{align}
    P_{A,t} (\mc W_t, \mc V_t)
     = \sum_{f \in \F,b \in \mc B} \delta_b \|\mb v_{f,b,t}\|^2
     + \sum_{f \in \F} \beta \| \mb w_{f,t}\|^2,
  \end{align}
where $\delta_b$ and $\beta$ are slopes of the load-dependent power constant for SBS $b$ and CP, respectively. 
{Preliminary studies \cite{tao2016content,peng2017layered} generally considered uncoded caching and delivered individual contents to each SBS from the CP via unicast. In this work, the employment of MDS codes can enable fronthaul multicast by transmitting coded contents to associated SBSs using the same communication resource. Therefore, it helps to reduce traffic load and power consumption \cite{liao2017coding}.} 

\section{Problem Formulation and Decomposition}
In this section, we first formulate the mixed timescale problem, then investigate the problem decomposition for the two-stage updating scheme, and finally identify the main challenges for algorithm design.

\subsection{Mixed Timescale Problem Formulation}
In this work, our task is to determine cache allocation matrix $\mb L$, and transmission policy $\Psi_t = \{\mc W_t,\mc V_t, \mb E_t\}$, with the goal of minimizing content delivery power consumption $P_{A,t} (\mc W_t, \mc V_t)$ during the next $T$ time frames while providing QoS guarantees for users.
By recalling Proposition 1 and \eqref{eq:mulrat}, we observe that the multicast beamformers $\mc W = \{\mc W_t, \forall t \in
\mc T\}$ and $\mc V = \{ \mc V_t, \forall t \in \mc T\}$ also depend on cache allocation $\mb L$ and SBS clustering $\mc E = \{\mb E_t, \forall t \in \mc T\}$, with $\mc T = \{t_0+1, \cdots, t_0+T\}$ and $t_0$ being the current cache-updating frame. Hence, it is necessary to jointly optimize the large timescale variable $\mb L$ together with the small timescale variables $\Psi = \{\Psi_t, t \in \mc T\}$, which yields the following mixed timescale power minimization problem:
\begin{subequations} \label{eq:main_problem}
  \begin{align}
    \min_{\mb L, \Psi}~&
    \textstyle\sum_{t \in \mc T}  P_{A,t} (\mc W_t, \mc V_t), \\
    s.t. ~&  \textstyle\sum_{f \in \mc F} l_{f,b}s_f \leq S_b, ~ b \in \B, \label{b}\\
    & 0 \leq l_{f,b} \leq 1, ~\forall f \in \mc F, b \in \mc B, \label{c}\\
    &   \mb E_t \in \{0,1\} ^{F_{\text{req,t}} \times B}, \forall t~\in \mc T, \label{eq:binary}\\
    &  \eqref{p1d}, \eqref{p1c}, \eqref{p1b}, \eqref{eq:FH}, \forall t~\in \mc T \label{d},
\end{align}
\end{subequations}
where
constraint \eqref{b} is due to the storage limit at each SBS; and constraints \eqref{p1b} and \eqref{eq:FH} are used to provide QoS guarantees for content download. It should be noted that the problem in \eqref{eq:main_problem} is {\it anti-causal} at the current cache-updating frame $t_0$ because the users' requested contents $\F, t\in \mc T$, and their CSI in the next $T$ frames are generally unavailable. {Notably, problem \eqref{eq:main_problem} may not be feasible if the SINR threshold $\{\gamma_f\}$ in \eqref{p1b} is too large or the CSI of users in different multicast groups are strongly correlated \cite{tao2016content,xiang2012coordinated}. Readers are referred to \cite{xiang2012coordinated} for a feasibility condition of multicast beamforming optimization under QoS guarantee constraints in multicell networks. Herein, we elaborate on addressing problem \eqref{eq:main_problem} when it is feasible.}

\subsection{Practical Implementation for Two-Stage Updating Scheme}
{To circumvent the difficulty of anti-causality of the problem in~\eqref{eq:main_problem}, we observe that, in practice, 
{channel statistics are approximately the same for two consecutive online content delivery frames.} Therefore, by appropriately making use of 
historical requests up to the current cache-updating frame $t_0$, we may obtain an accurate prediction of caching contents for future requests. Based on the above considerations, we aim to propose a two-stage updating scheme to handle the problem in~\eqref{eq:main_problem}.

For unified implementation, we consider that the CP operates in transmission block fashion. As depicted in Fig.~\ref{JointScheme}, each transmission block consists of $T$ frames, and all SBSs' caches are updated by the end of each transmission block\footnote{In practice, the CP may adjust the cache-updating frames by taking into account traffic load \cite{bharath2016learning}.}. {Moreover, $T$ should be sufficiently large to aggregate enough historical observations for cache updating, i.e., $T \gg 1$.}
\begin{figure}[h]
	\centering
	\includegraphics[scale=0.63]{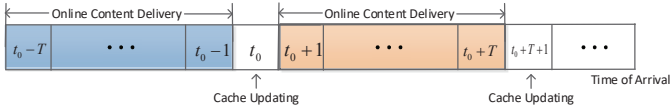}
	\caption{A mixed timescale model for periodic cache updating and short-term content delivery. }
	\label{JointScheme}
\end{figure}
 In the first stage, we update the cache allocation $\mathbf L$ (together with the clustering and beamforming) by using the {\it historical} content requests and CSI in the past $T$ time frames, i.e., $\mc T' = \{t_0 - T, t_0 - T + 1, \ldots, t_0 -1\}.$ In the second short-term delivery stage, we fix the cache allocation $\mathbf L$ based on the results of the first stage, and optimize transmission policy $\Psi_t$ according to the current CSI and content requests $\F$, for $ \forall t\in \mc T$.

\subsubsection{Long-Term Cache Updating Design} At the cache-updating frame $t_0$ (e.g., the end of each transmission block), the historical content requests and CSI are already available in the CP. Therefore, cache allocation matrix $\mb L$ is given by solving the following problem $\mc P$.
  \begin{align}
    \mc P({\mb L, \Psi}) : \min_{\mb L, \Psi}~~&
    \textstyle\sum_{t \in \mc T'} P_{A,t}, ~
    {\rm s.t.} ~ \eqref{b} - \eqref{d}, \label{pro:LTC}
\end{align}
{where we recall $\mc T' = \{t_0 - T, \ldots, t_0 -1 \}$ in $\mc P$.} Subsequently, solution $\mb L^*$ to problem $\mc P$ is used to update the cached contents, which will serve for content delivery in the next block $\mc T$.
\subsubsection{Short-Term Content Delivery Design} This is implemented
at each frame $t \in \mc T$.
The CP needs to schedule SBS clustering and beamforming strategy to satisfy users' requests $\pi_t$, given the cached contents in SBSs $\mb L^*$ and instantaneous CSI. Accordingly, transmission policy $\Psi_t$ is optimized by solving the following problem.
\begin{align}
   \mc P_t (\Psi_t|\mb L^*):\min_{ \Psi_t}~& P_{A,t}(\mc W_t, \mc V_t) \\
   ~~~{\rm s.t.}~& \mb E_t \in \{0,1\} ^{F_{\text{req,t}} \times B}, \label{eq:bint}\\
   				&\eqref{p1d}, \eqref{p1c}, \eqref{p1b}, \eqref{eq:FH},
\end{align}
for $t=t_0+1, \ldots, t_0 +T$.

\subsection{Problem Analysis and Reformulation}
For ease of discussion, we first analyze the short-term content delivery design. Problem $\mc P_t$ incorporates beamforming design for both edge and fronthaul links, as well as SBS clustering, which is difficult to tackle. The main challenges are identified as follows:

$\bullet$ {\it Binary Variables for SBS Clustering:} Due to the involvement of binary variables $\mb E_t$, $\mc P_t$ is an MINLP. Hence, attaining an optimal solution for $\mc P_t$ is NP-hard, in general. To address the discontinuity in $\mc P_t$, we reformulate the binary constraint \eqref{eq:bint} into $e_{f,b,t}^2  -e_{f,b,t} = 0$,
which can be further written as two continuous constraints
\begin{align}
  e_{f,b,t}^2 - e_{f,b,t} \geq 0,  \forall f \in \F, b \in \mc B, \label{h}\\
  e_{f,b,t}^2 - e_{f,b,t} \leq 0, \forall f \in \F, b \in \mc B. \label{i}
\end{align}
The equilibrium constraint \eqref{p1c} can also be equivalently written as a convex constraint
\begin{align}
  \|\mb v_{f,b,t}\|_2 \leq e_{f,b,t}\sqrt{P_b}, \forall f \in \F, b \in \mc B.  \label{m}
\end{align}
It can be verified that the associated beamformer $\mb v_{f,b,t} = 0$ for $e_{f,b,t} = 0$.

$\bullet$ {\it Nonconvexity of SINR Constraints \eqref{p1b}:} Constraint \eqref{p1b} can be transformed as
\begin{align}
   \gamma_{f}  g_{k,t}({\mc V_t}) - |\mb h_{k,t}^H \mb v_{f,t}|^2 \leq 0, \forall k \in \mc G_{f,t}, \forall f \in \F, \label{l}
\end{align}
where the function on the left-hand side is the difference of two convex (DC) functions. Hence, to take advantage of this structure, we can adopt the CCCP technique to efficiently address this challenge.

$\bullet$ {\it Nonconvexity of Fronthaul Multicast Data Rate \eqref{eq:mulrat}:} The multicast rate \eqref{eq:mulrat} for the fronthaul link is yielded by applying MDS coded caching to our studied model. We can calculate the fronthaul power consumption according to multicast beamformers from the CP. {To make \eqref{eq:mulrat} tractable, we consider the following constraint:
\begin{align}
    2^{\frac{\FH + \tau_0 (e_{f,b,t} - 1)}{B_2}} - {\|\mb H_{b,t}^H \mb w_{f,t}\|^2}/{z_b^2} \leq 1 , \forall b \in \mc B, f \in \F, \label{Prop2}
\end{align}
where parameter $\tau_0 \geq \max_{f \in \mc F_{\req,t}} R_f$, which is a sufficiently large constant.} 


$\bullet$ {\it Nonconvexity of Fronthaul Constraints \eqref{eq:FH}:} To avoid download interruptions and provide better QoS guarantees, the edge transmission rate should be restricted by the fronthaul capacity \eqref{eq:FH}, which is also a nonconvex constraint. To make constraint \eqref{eq:FH} more tractable, it can be re-expressed as
\begin{align}
  R_{f,t}^{\text{FH}} \geq (1 - l_{f,b}) e_{f,b,t}R_f, \forall f \in \F, b\in \mc B. \label{eq:FH2}
\end{align}

According to the above analysis, eventually, problem $\mc P_t$ can be reformulated as
        \begin{align}
            \mc P_{1,t} ( \Psi_t, R_{f,t}^{\text{FH}}| \mb L^*):\min_{\Psi_t,R_{f,t}^{\text{FH}}}~&  {P_t(\mc W_t,\mc V_t)} \\
             {\rm s.t.} ~& \eqref{p1d}, \eqref{h} - \eqref{eq:FH2} \label{eq:P1t}.
        \end{align}
\vspace*{0.2cm}\\
{\it {\bf Proposition 2:} Problem $P_{1,t}( \Psi_t, R_{f,t}^{\text{FH}}| \mb L^*)$ attains the same optimal value as that of problem $\mc P_t (\Psi_t|\mb L^*)$.     
\begin{proof}
See Appendix B. \hfill $\blacksquare$
\end{proof}}
As can be observed in problem $\mc P_{1,t}$, constraints \eqref{h}, \eqref{l} and \eqref{Prop2}, which account for the coupling relations among binary variables for SBS clustering, edge and fronthaul beamformers, constitute a main hinderance to solve the proposed problem. 
Indeed, developing an efficient algorithm is very challenging.

\section{Proposed Algorithm for Short-Term Content Delivery Design $\mc P_{1,t}$}

In this section, we focus on developing an efficient algorithm for short-term content delivery design $\mc P_{1,t}$. To deal with the aforementioned challenges, we first leverage a Frobenius-Norm penalty approach to address the difficulty of binary variables, and then propose a CCCP-based algorithm with convergence guarantee.

\subsection{A Frobenius-Norm Penalty Approach}
Recall that the binary constraint of SBS clustering variable $e_{f,b, t}$ has been converted into continuous constraints \eqref{h} and \eqref{i}. However, the tight coupling of these two constraints may make it difficult to move away from the initial point in algorithm implementation.

To cope with this difficulty, inspired by the idea in \cite{lipp2016variations}, some nonnegative slack variables $\mb E_t' = [e_{f,b,t}']$ can be employed to relax constraint \eqref{h} into
\begin{align}
  e_{f,b,t} - e_{f,b,t}^2 & \leq e_{f,b,t}', \forall f \in \F, b \in \mc B, \label{t}
\end{align}
where $e_{f,b,t}' \geq 0$; and then, the violation of constraint \eqref{t} can be penalized by minimizing the Frobenius Norm of the slack variable $\mb E_t'$. Accordingly, problem $\mc P_{1,t}$ can be transformed as
\begin{subequations}
        \begin{align}
            \mc R_{1,t} ( \Theta_t | \mb L^*): \min_{ \Theta_t}~&  {P_t(\mc W_t,\mc V_t)}+ \lambda\|\mb E_t'\|_F, \label{r0a}\\
            {\rm s.t.} ~& e_{f,b,t}' \geq 0, \forall f \in \F, b \in \mc B,  \label{r0b}\\
            & \eqref{p1d}, \eqref{i} - \eqref{eq:FH2}, \eqref{t},  \label{r0c}
        \end{align}
\end{subequations}
where variables $\Theta_t = \{\Psi_t, R_{f,t}^{\text{FH}}, \mb E_t'\}$; and penalty parameter $\lambda \geq 0$. 
It is worth noting that although the original MINLP $\mc P_t$ is reformulated into a tractable form, obtaining the optimal solution of the original MINLP $\mc P_t$ remains hard since the reformulated problem $\mc R_{1,t}$ is also nonconvex.

\subsection{Proposed Algorithm for Short-Term Content Delivery}
Problem $\mc R_{1,t}$ contains DC constraints \eqref{l}, \eqref{Prop2} and
\eqref{t}, and other convex constraints and objective function. Thus, problem $\mc R_{1,t}$ takes the following general form of DC program,
\begin{subequations}
\label{pro:DC}
\begin{align}
  \min_{\boldsymbol\theta} ~~&f_0 (\boldsymbol\theta) - r_0 (\boldsymbol\theta) \\
  {\rm s.t.}~~& f_n (\boldsymbol\theta) - r_n (\boldsymbol\theta) \leq 0, ~~ n = 1,2, \cdots, n_0,
\end{align}
\end{subequations}
where functions $f_n(\cdot)$ and $r_n (\cdot)$ are convex. When $g_n (\cdot)$ is not linear, the DC program is not convex and it is generally difficult to obtain an optimal solution. An efficient approach for solving a DC program is to utilize the  CCCP technique, which may attain a local optimal solution \cite{lipp2016variations}. In particular, $\boldsymbol\theta$ is updated by solving the following convex problem
\begin{subequations}
\label{pro:CCCP}
\begin{align}
  \min_{\boldsymbol\theta} ~~&f_0 (\boldsymbol\theta) - \nabla r_0 (\boldsymbol\theta ^{(i)})^T \boldsymbol\theta  \\
  {\rm s.t.}~~& f_n (\boldsymbol\theta) - \big[ r_n (\boldsymbol\theta ^{(i)}) + \nabla r_n (\boldsymbol\theta ^{(i)})^T (\boldsymbol\theta - \boldsymbol\theta^{(i)})\big]  \leq 0, ~~ \forall n,
\end{align}
\end{subequations}
where parameter $\boldsymbol\theta ^{(i)}$ is an optimal solution in the $i$th iteration of CCCP. 

Therefore, we handle problem $\mc R_{1,t} ( \Theta_t | \mb L^*)$ based on the above-mentioned principles. Firstly, we define the following functions 
	$r_{1,k} (\mb v_{f,t}) = |\mb h_{k,t}^H \mb v_{f,t}|^2, \forall k$, $
	r_{2,b} (\mb w_{f,t}) = {\|\mb H_{b,t}^H \mb w_{f,t}\|^2}/{z_b^2}$ $\forall b,$ and $
	r_3 (e_{f,b,t}) = e_{f,b,t}^2.$ 
Consequently, by using the first-order Taylor expansion, the lower-bounds for these functions are given by
\begin{align}
	\!\hat r_{1,k} (\mb v_{f,t}) & = 2\text{Re} \{ \mb v_{f,t}^{(i)H} \mb h_{k,t} \mb h_{k,t}^H \mb v_{f,t}\} - \big|{\mb h_{k,t}^H \mb v_{f,t}^{(i)}}\big|^2, \forall k \in \mc G_{f,t},\\
	\hat r_{2,b} (\mb w_{f,t}) & = 2\text{Re} \{ \mb w_{f,t}^{(i)H} \mb H_{b,t} \mb H_{b,t}^H \mb w_{f,t}\}/{z_b^2} - \big\|{\mb H_{b,t}^H \mb w_{f,t}^{(i)}}\big\|^2/{z_b^2}, \\
	\hat r_3 (e_{f,b,t}) &= (2 e_{f,b,t}^{(i )} -1 )e_{f,b,t},
\end{align}
for any $f,b,t$, respectively, where $\Psi_t^{(i)} = \{\mc W_t^{(i)}, \mc V_t^{(i)}, \mb E_t^{(i)}\}$ denotes the solution in the $i$-th iteration. Accordingly, problem $\mc R_{1,t}$ can be addressed by successively solving the following approximate problem $\mc R_{2,t} ( \Theta_t|\Psi_t^{(i)}, \mb L^*)$:
\begin{subequations}
\label{pro:r2t}
        \begin{align}
            \!\min_{ \Theta_t}~&  {P_t(\mc W_t,\mc V_t)}+ \lambda\|\mb E_t'\|_F \\
            {\rm s.t.} ~&  \gamma_{f,t}g_{k,t}(\mc V_t) - \hat r_{1,k} (\mb v_{f,t}) \leq 0,\forall k \in \mc G_{f,t}, \forall f \in \F, \\
            & 2^{\frac{\FH + \tau_0 (e_{f,b,t} - 1)}{B_2}} - \hat r_{2,b} (\mb w_{f,t}) \leq 1 , \forall f \in \F, b \in \mc B,  \label{eq:exp}\\
            & (e_{f,b,t}^{(i)})^2  -\hat r_3 (e_{f,b,t})  \leq e_{f,b,t}', \forall f \in \F, b \in \mc B,  \\
            & \eqref{p1d}, \eqref{i}, \eqref{m}, \eqref{eq:FH2}, \eqref{r0b}.
        \end{align}
\end{subequations}
Problem \eqref{pro:r2t} is convex, and can be efficiently solved by interior point methods (IPM) in polynomial times using a standard solver, such as CVX \cite{grant2008cvx}.

\begin{algorithm}[h]
\label{AL1}
\caption{Proposed Algorithm for Short-Term Content Delivery Design $\mc P_{1,t}$}
\begin{algorithmic}[1]
\State  {\bf Initialize} $i =0$ , $ \mb E_t^{(0)}$, $\lambda > 0, \tau >1, \lambda_{\max}, t, \mb L^*$
\State  Find a feasible point for $\mc W_t^{(0)}$ and $\mc V_t^{(0)}$
\State  {\bf Repeat}
\State ~~~~Solve the convex problem $\mc R_{2,t}( \Theta_t|\Psi_t^{(i)}, \mb L^*) $ by CVX and obtain $\Psi_t^{(i+1)}$
\State ~~~~Update $\lambda \leftarrow \min\{\lambda_{\max}, \tau\lambda\}$
\State ~~~~Update $i \leftarrow i+1$
\State {\bf Until} some stopping criterion is satisfied
\State {\bf Output} transmission policy $\widehat{\Psi}_t = \{\widehat{\mc W_t}, \widehat{\mc V_t}, \widehat{\mb E}_t\}$
\end{algorithmic}
\end{algorithm}
The corresponding implementation procedure is summarized in Algorithm 1. To obtain a good initial point for nonconvex problem $\mc R_{1,t}$, penalty parameter $\lambda$ is initialized with a small value $\lambda_0 >0 $ and gradually increased to force slack variable $e_{f,b,t}'$ to approach 0.} It is easy to verify that each variable $e_{f,b,t}$ attains a binary value if slack variable $e_{f,b,t}' = 0$.
In practical realization, parameter $\lambda_{\max}$ is generally set to be sufficiently large \cite{lipp2016variations}. 
The sequence of the objective value of problem $\mc R_0$, generated by Algorithm 1 during the iterative process, will converge based on the principle of CCCP \cite{lipp2016variations}. {Furthermore, we consider a small threshold $\epsilon > 0$ for projecting $\mb E_t$ into a feasible solution if the slack variables are not exactly zero. Namely, when the resulting $e_{f,b,t} < \epsilon$, we let $e_{f,b,t} = 0$, otherwise $e_{f,b,t} = 1$; and the other decision variables need to be optimized accordingly.}
Note that, solving optimization problem $\mc R_{2,t}$ in step 4 dominates the overall computational complexity. 
{Specifically, we adopt the fundamentals of complexity analysis for IPM \cite{boyd2004convex}. In particular, constraint \eqref{eq:exp} involves the exponential operation, which is usually handled by the first-order Taylor expansion \cite{grant2008cvx}; and other quadratic constraints can be converted into linear matrix inequality (LMI) constraints. This is a standard approach. Readers are referred to \cite{hindi2004tutorial,boyd2004convex} in greater detail. For notational convenience, the cardinality of $\mc F_{\req,t}$ is denoted by $F_0$. 
As such, solving problem $\mc R_{2,t}$ involves $K_t$ LMI constraints of dimension $BMF_0+1$, $B$ LMI constraints of dimension $MF_0 +1$, $BF_0$ second-order cone (SOC) constraints of dimension $M+1$. Eventually, the  complexity of Algorithm 1 is in the order of
{\small
$
	\sqrt{K_tX + B(MF_0 +1) +2BF_0} n \big[ K_tX^3 + B(MF_0+1)^3 + n K_tX^2 + nB(MF_0 + 1)^2 + BF_0(M+1)^2 +n^2\big], 
$
}
where $X = BMF_0+1$, and $n$ denotes the dimension of decision variables\footnote{Low dimensional constraints are omitted which shall not impact complexity order of the whole algorithm.}, i.e., $F_0(1+2B+N+MB)$. As a comparison, a sparsity-based approach adopted in \cite{peng2017layered} for beamformer design in cache-enabled system is as follows: each binary variable $e_{f,b,t}$ is first replaced by $\|\|\mb v_{f,b,t}\|^2\|_0$, and then the original problem becomes a sparse optimization problem, which is further approximated by using the $l_1/l_2$-norm. However, this approach shall suffer from high computational complexity because it comes at cost of lifting the size of problem. Moreover, applying this approach may turn problem $\mc P_{1,t}$ to be more complicated than the original one due to the coupling constraints \eqref{Prop2} and \eqref{eq:FH2}.   
}

\section{Proposed Two-Stage Content Caching Framework}
In this section, we first present the proposed algorithm for long-term cache updating design, then develop another algorithm to reduce computational complexity, and finally introduce a practical implementation of the proposed two-stage content caching framework.

\subsection{Proposed Algorithm for Long-Term Cache Updating Design $\mc P(\mb L, \Psi)$}
We observe that when the cache allocation matrix $\mb L$ is fixed, problem $\mc P$ can be decomposed into a group of independent subproblems $\mc P_{1,t} ( \Psi_t, R_{f,t}^{\text{FH}}| \mb L), {\forall t \in \mc T'}$. Hence, to take advantage of this beneficial property, we leverage the alternating  method to solve problem $\mc P$. Specifically, by fixing $\mb L = \mb L^{(i)}$, the first block $\Psi$ can be updated by solving a group of subproblems $\mc P_{1,t} (\Psi_t, R_{f,t}^{\text{FH}} |\mb L^{(i)}), {\forall t \in \mc T'} $ in parallel. {For the other block $\mb L$, it can be updated by checking the feasibility of problem $\mc P$ by fixing $\Psi = \{\mc W^{(i)}, \mc V^{(i)}, \mc E^{(i)}\}$ since the objective function in $\mc P$ is independent of $\mb L$.  
Note that by recalling Proposition 1 and \eqref{eq:mulrat}, one can observe that a smaller fronthaul data rate, i.e., lower-bounded by the right-hand side of \eqref{eq:FH}, is always likely to result in lower fronthaul power consumption.
Accordingly, an efficient method to update $\mb L$ is given by solving the following problem as 
\begin{align}
 \mc P_2: \min_{\mb L}~ &\sum_{t \in \mc T'}  \sum_{f \in \F} \max_{b \in \mc B}~ (1 - l_{f,b}) e_{f,b,t}^{(i)}R_f
\\
 {\rm s.t.}~~
 &\eqref{b}, \eqref{c},
\end{align}
which is a convex optimization with linear constraints, which can also be easily solved by IPM. 
Towards this end, the entire process for addressing long-term cache updating design is summarized in Algorithm 2. {Specifically, an inexact block coordinate descent (BCD) approach is adopted because the resulting subproblem $\mc P_{1,t}$ is still nonconvex. Thus, Algorithm 2 can generate a decreasing sequence of objective values, which is lower bounded and able to converge due to the monotone convergence theorem; however, owing to the nonconvexity of this problem, a suboptimal solution may be attained \cite{razaviyayn2014parallel}.}

\begin{algorithm}[h]
\caption{Proposed Algorithm for Periodic Cache Updating Design (PCUD) $\mc P(\mb L, \Psi)$}
\begin{algorithmic}[1]\label{AL1}
\State {\bf Initialize} $i =0$ , $ \mb L^{(0)}$
\Repeat
\State Fixing $\mb L = \mb L^{(i)}$, solve problems $\mc P_{1,t} (\Psi_t, R_{f,t}^{\text{FH}} |\mb L^{(i)})$ for ${\forall t \in \mc T'}$ in parallel by Algorithm 1, and the solution is defined as $\Psi_t^{(i +1)}$
\State Fixing $\Psi_t = \Psi_t^{(i +1)}$, solve problem $\mc P_2$ and the solution is defined as $\mb L^{(i + 1)}$
\State  Update $i \leftarrow i+1$
\Until {some stopping criterion is satisfied}
\State {\bf Output:} $\mb L$ for the next block $\mc T$
\end{algorithmic}
\end{algorithm}
Obviously, finding an effective cache-updating policy comes at the cost of re-optimization of the beamformers from all SBSs and the CP, as well as the SBS clustering matrix for each frame $t \in \mc T'$. {Although the parallel computing in step 3 helps to decrease the execution time, the computational complexity of this algorithm is in the order $\mc O(T)$ of that of Algorithm 1.}
From the perspective of practical implementation, a compromise between computational complexity and network performance should be considered. 

\subsection{A Learning-Based Approach for Cache Updating Design}
{Recall that, at each frame $t\in \mc T'$, the CP needs to calculate SBS cluster $\mb E_t$ by addressing $\mc P_{1,t}$. Thus, at cache updating frame $t_0$, the CP has known historical SBS clustering policies $\mc E' = \{\mb E_t, \forall t \in \mc T'\}$. Hereafter, we propose a learning-based method to perform cache updating by properly utilizing historical observations $\{\pi_t, \forall t \in \mc T'\}$ and SBS clustering $\mc E'$.} 

In some previous studies, such as \cite{bharath2016learning}, the distribution of content preference is globally estimated and then used to optimize cache resources at all SBSs homogeneously. Nevertheless, in practice, the same content may receive different attention from users served by different SBSs. Accordingly, we capture content preference by considering users' local behavior towards content requests. 

In particular, aiming at learning the local content preference, the CP calculates the $F \times 1$-vector $\mb q_{b}'= [q_{f,b}']$ for any $b$ at the end of revealed block $\mc T'$. Specifically, each element $q_{f,b}'$ implies the frequency of content $f$ being served by SBS $b$ in $\mc T'$, i.e.,
\begin{align}
	q_{f,b}' = \frac{ \sum_{t \in \mc T'}{N}'_{f,t} e_{f,b,t}}{\sum_{t \in \mc T', f \in \mc F}{N}'_{f,t} e_{f,b,t}}, \label{EstPre}
\end{align}
under SBS cooperation, where ${N}_{f,t}'$ denotes the number of received requests for content $f$ at frame $t$. Clearly, $\sum_{f \in \mc F} q_{f,b}' = 1, \forall b$. 	
Moreover, to reduce power consumption, the caching contents are envisioned to be frequently reused as so to alleviate duplicated fronthaul transmissions. Hence, instead of solving $\mc P(\mb L, \Psi)$, a cache updating problem is formulated by minimizing the total fronthaul rate, i.e., 
\begin{align}
	\mc P_3: \min_{\mb L} ~&\sum_{f \in \mc F} \max_{b \in \mc B}~q_{f,b}'(1 - l_{f,b}) R_f\\
	{\rm s.t.} ~ & \eqref{b}, \eqref{c}.
\end{align}
Note that, the utilization of the local content preference $\{q'_{f,b}\}$ in $\mc P_3$ allows the CP to update $\mb L$ by tracking and adapting to historical patterns of SBS collaboration and users' requests. One can easily verify that problem $\mc P_3$ is convex, which can be efficiently solved by IPM. The detailed process for the proposed low-complexity cache updating design is summarized in Algorithm 3. 
\begin{algorithm}[h]
\caption{Proposed Low-Complexity Algorithm for Periodic Cache Updating Design (LC-PCUD) {$\mc P_3$}}
\begin{algorithmic}[1]\label{LT2}
\State {\bf Initialize} $t = t_0$, $\mc T' = \{t_0 - T, \cdots, t_0 -1\}$
\State Calculate the number of received requests $N_{f,t}'$ for file $f$, $\forall t \in \mc T'$
\State Estimate local content preference $\{q'_{f,b}, \forall f, b\}$ from the historical observations and SBS clustering $\mc E'$ through \eqref{EstPre}
\State Obtain the prediction of cache updating matrix $\mb L$ by solving problem $\mc P_3$
\State {\bf Output:} $\mb L$ for the next block $\mc T$
\end{algorithmic}
\end{algorithm}

\begin{figure}[h]
	\centering
	\includegraphics[scale=0.58]{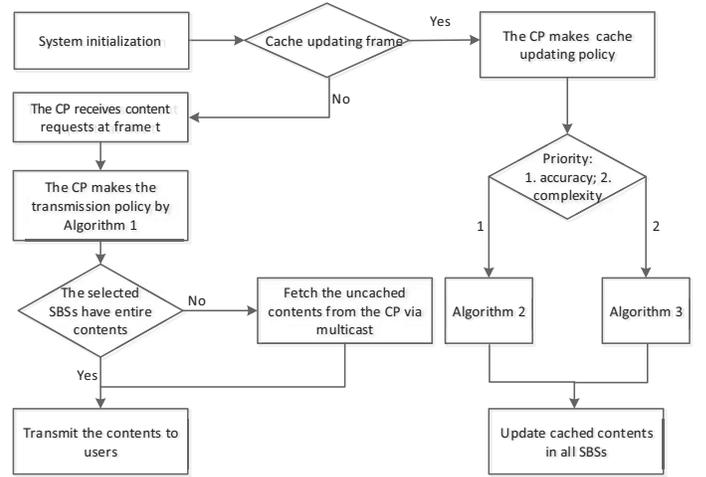}
	\caption{Brief process of the two-stage content caching framework for long-term cache updating and short-term content delivery in a cache-enabled C-SCNs. }
	\label{PropFram}
\end{figure}

\subsection{Practical Implementation of the Two-Stage Content Caching Framework}
{A general flowchart is presented in Fig. \ref{PropFram}, which outlines the practical implementation of the proposed two-stage content caching framework. Specifically, in the short-term delivery stage, the CP runs Algorithm 1 to make the short-term transmission policy for each frame $t$, i.e., $\{\mc W_t, \mc V_t, \mb E_t\}$, based on the instantaneous requests $\pi_t$ and CSI, as well as the cache status in the local SBSs. Accordingly, the CP delivers the missed MDS coded parity bits to the selected SBSs via multicast transmissions while each SBS cluster provides download service with guaranteed QoS to the associated multicast group by using cooperative beamforming. With regard to long-term cache-updating stage, the cached contents in all SBSs are updated by the end of each transmission block that contains $T \gg 1$ frames.
The CP coordinates the cache updating policy by leveraging the historical observations and the collaborative patterns of SBSs. From the perspective of practical implementation, the CP can run either Algorithm 2 or 3 so as to achieve a good accuracy-complexity trade-off.}

\section{Performance Evaluation}
Consider a cache-enabled C-SCNs that covers a hexagonal cell with an edge length of 500 m. The CP is located at the center of this area, which coordinates five SBSs that are randomly deployed within this cell. The CP is equipped with eight antennas, and each SBS is equipped with four antennas. Consider that all SBSs have an equivalent fractional caching capacity $\mu$, where $\mu = \sum_{b \in \B}S_b/(B\sum_{f \in \mc F} s_f)$. In addition, several single-antenna users are randomly distributed within the cell, except for the area in a radius of 30 m around SBSs and the CP. The channel bandwidth of edge links is 10 MHz.
With regard to wireless channel, we consider that: the path-loss model $PL(\text{dB}) = 148.1 + 37.6\lg(d)$; $d (\text {km})$ denotes the distance between each SBS and each user (or the CP) \cite{tao2016content}; the antenna power gain is 10 dBi; and the log-normal shadowing parameter is $8$ dB; and the small scale fading is modeled as Rayleigh fading with zero mean and unit variance that is i.i.d. at each frame. We also consider that all users are quasi-static. In addition, there are 100 contents in the library. To be realistic, we consider that users have different preference patterns towards these contents. Specifically, a batch of four users share a unique Zipf distribution, which is specified by the popularity rank order of all contents and a skewness factor \cite{liao2017coding}.
The default settings are: the fractional caching capacity in each SBS is 20$\%$; the maximum transmit power for each SBS is 0 dB; the SINR requirement of each multicast group is 10 dB; {users' requests happen 100 times in each transmission block}; a total of three preference patterns towards contents are considered and the users in each preference pattern are randomly active at each frame; for each preference pattern, the popularity rank order of all contents is randomly generated and the skewness factor is also randomly picked from region $[1,3]$, similar to \cite{liao2017coding}; the total fronthaul bandwidth is 60 MHz, and we allocate 5 MHz to serve content delivery for each multicast group through fronthaul (unless stated otherwise). With regard to power consumption model, we consider typical values $\delta_b = 2.7$ and $\beta = 4.0$ \cite{auer2011much}.

We first illustrate the convergence behavior of the proposed penalty-based algorithm for short-term content delivery design.
As depicted in Fig. \ref{Fig:AlgCon}, five simulation trials are conducted. Specifically, in each trial, we initialize penalty parameter $\lambda$ with $\lambda_0 = 1$ and increase it by three every iteration until $\lambda$ reaches the maximum value $\lambda_{\max} = 50$; meanwhile, each element of $\mb E_t^{(0)}$ is randomly generated within $[0,1]$, but prefixed as the same value for each trial. Moreover, a small threshold $\epsilon = 0.01$ is used for quantization. 
For comparison, in each trial, the beamformers from each SBS and the CP are randomly and independently initialized as any feasible point. It can be observed that {all curves descend firstly and then are slightly lifted due to the increase of the penalty term in the objective function.} After four iterations, the penalty parameter $\lambda$ reaches its maximum value, and then the objective value starts decreasing. Furthermore, we can find that the objective values of all trials almost converge to the same value in 5 - 10 iterations. This finding implies that the proposed algorithm converges very fast and is not sensitive to the initial points of beamformers, which is suitable for practical implementation.
\begin{figure}[h]
  \centering
  \includegraphics[scale=0.39]{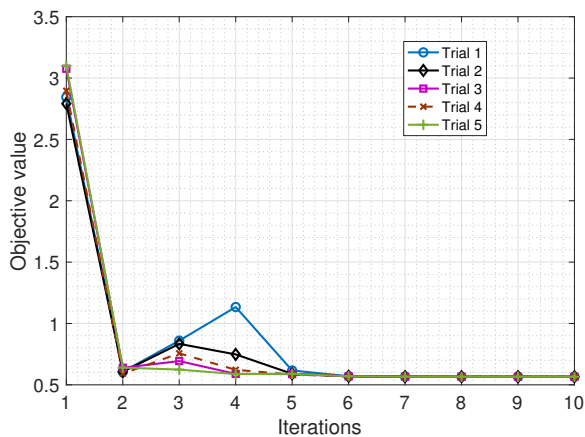}
  \caption{Convergence behavior of Algorithm 1.}
  \label{Fig:AlgCon}
\end{figure}
Next, we evaluate the caching performance of the proposed two-stage framework under different network settings. In particular, the following baselines are considered:
\begin{itemize}
	\item {\bf Uniform Caching (UC):} Each SBS randomly and uniformly fetches the segments of each content, i.e., $l_{f,b} = \mu, \forall f \in \mc F, b \in \mc B$. This baseline works as an upper bound.
	\item {\bf Least Recently Used (LRU):} Like the idea in \cite{cao1997cost}, at each frame, each SBS stores the newly arrived contents according to users' requests. If overflow error occurs, the least recently used contents will be removed.
	\item {\bf Genie Aided Caching (GAC):} By assuming that we have already acquired the genie information, i.e., users' requests $\{\pi_t, t \in \mc T\}$ and CSI in the next transmission block, the cached contents are updated by Algorithm 2. This baseline works as a lower-bound.
\end{itemize}
Meanwhile, the associated short-term transmission policies of the above baselines are all scheduled by the proposed MDS codes-aided scheme (e.g., Algorithm 1). To illustrate the benefits of fronthaul multicast and cooperative beamforming, we also consider another baseline as follows.
\begin{itemize}
	\item {\bf Two-Stage Fronthaul Unicast Caching (TS-FUC):} This is also a two-stage caching framework by using historical observations of users' requests and CSI, similar to the proposed framework. However, the short-term delivery scheme is handled by the idea in existing studies \cite{peng2017layered,tao2016content}, i.e., any cache miss is dealt with by costly fronthaul unicast.
\end{itemize}

\subsection{Effects of the Fronthaul Bandwidth}
Fig. \ref{Fig:FHC} illustrates the bottleneck effects of fronthaul bandwidth on the performance of cache-enabled C-SCNs. {Fronthaul bandwidth for each multicast group varies from 4 MHz to 10MHz.} Clearly, the proposed two-stage framework significantly outperforms TS-FUC, UC and LRU with an average of $74.1\%$, $47.7\%$ and $28.4\%$ reduction in long-term power consumption, respectively.
In addition, the curve of Algorithm 3 (i.e., LC-PCUD) is close to that of Algorithm 2 (i.e., PCUD), which demonstrates the proposed framework is able to achieve a desirable trade-off between performance and complexity. It is worth noting that, under the low fronthaul bandwidth region, i.e., 4 MHz - 6 MHz, power consumption degrades rapidly, which implies that fronthaul power consumption may dominate system performance over this region. When the fronthaul bandwidth is larger than 7 MHz, all schemes experience almost no significant reduction in power consumption. This observation may indicate that, given a sufficiently large fronthaul bandwidth, power consumption for content delivery is primarily determined by edge transmissions. Furthermore, it can be observed the curves of the proposed algorithms gradually approach the curve of the GAC lower-bound when the fronthaul bandwidth becomes large. This finding indicates that the proposed framework could cache the contents effectively by reasonably adapting to the potential patterns of SBS cooperation and the underlying distributions of local users' content preferences.

\subsection{Effects of Content Popularity}
As shown in Fig. \ref{Fig:PP}, we investigate the effects of content popularity on long-term power consumption by varying the number of preference patterns. We can find that when the number of preference patterns becomes larger, the MNO needs to satisfy users' requests at a cost of higher power consumption. The reason for this finding is that when more preference patterns are considered, more users are involved and thus the MNO may receive more content requests. Moreover, the proposed two-stage framework is always superior to TS-FUC, UC, and LRU with an average power reduction of $92.4\%$, $26.8\%$ and $20.7\%$,
respectively. Notably, the gaps between the proposed framework and TS-FUC become increasingly large as the MNO receives more content requests. This observation indicates that using unicast transmissions in fronthaul may cause substantial energy costs, which prevents SBS cooperation from being efficient and degrades system performance. 
Moreover, we can observe that the gaps between the proposed framework and GAC scheme vanish when the number of preference patterns becomes large.

From the above analysis, we can conclude that the proposed framework exhibits superior performance gain over the baselines of TS-FUC, UC and LRU, and is also able to approach the performance of GAC under broad fronthaul bandwidths and heterogeneous content preference scenarios. In addition, the proposed low-complexity algorithm is capable of achieving a good approximation of Algorithm 2. In the remaining simulations, we focus on performance comparison between the proposed framework (implemented by Algorithm 2) and the UC upper-bound, as well as the GAC lower-bound.


\begin{figure}[h]
			  \centering
			  \includegraphics[scale=0.40]{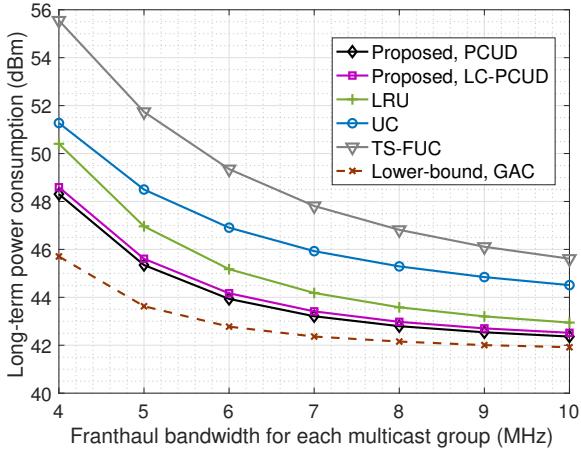}
			  \caption{The effects of fronthaul bandwidth.}
			  \label{Fig:FHC}
\end{figure}
\begin{figure}[h]
			  \centering
			  \includegraphics[scale=0.39]{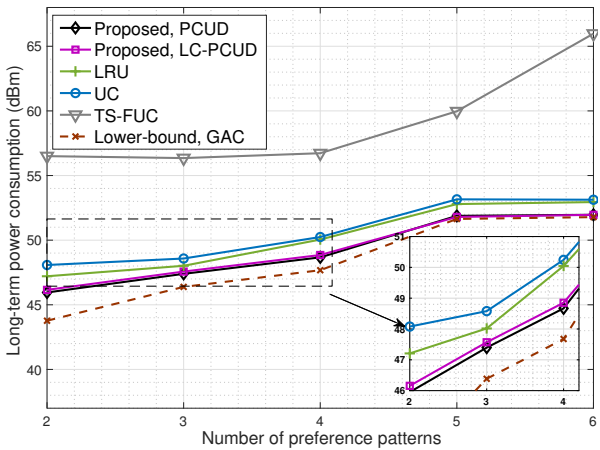}
			  \caption{The effects of the number of preference patterns.}
			  \label{Fig:PP}
\end{figure}
\begin{figure}[h]
  \begin{minipage}{0.5\linewidth}
   \centering
  \includegraphics[scale=0.39]{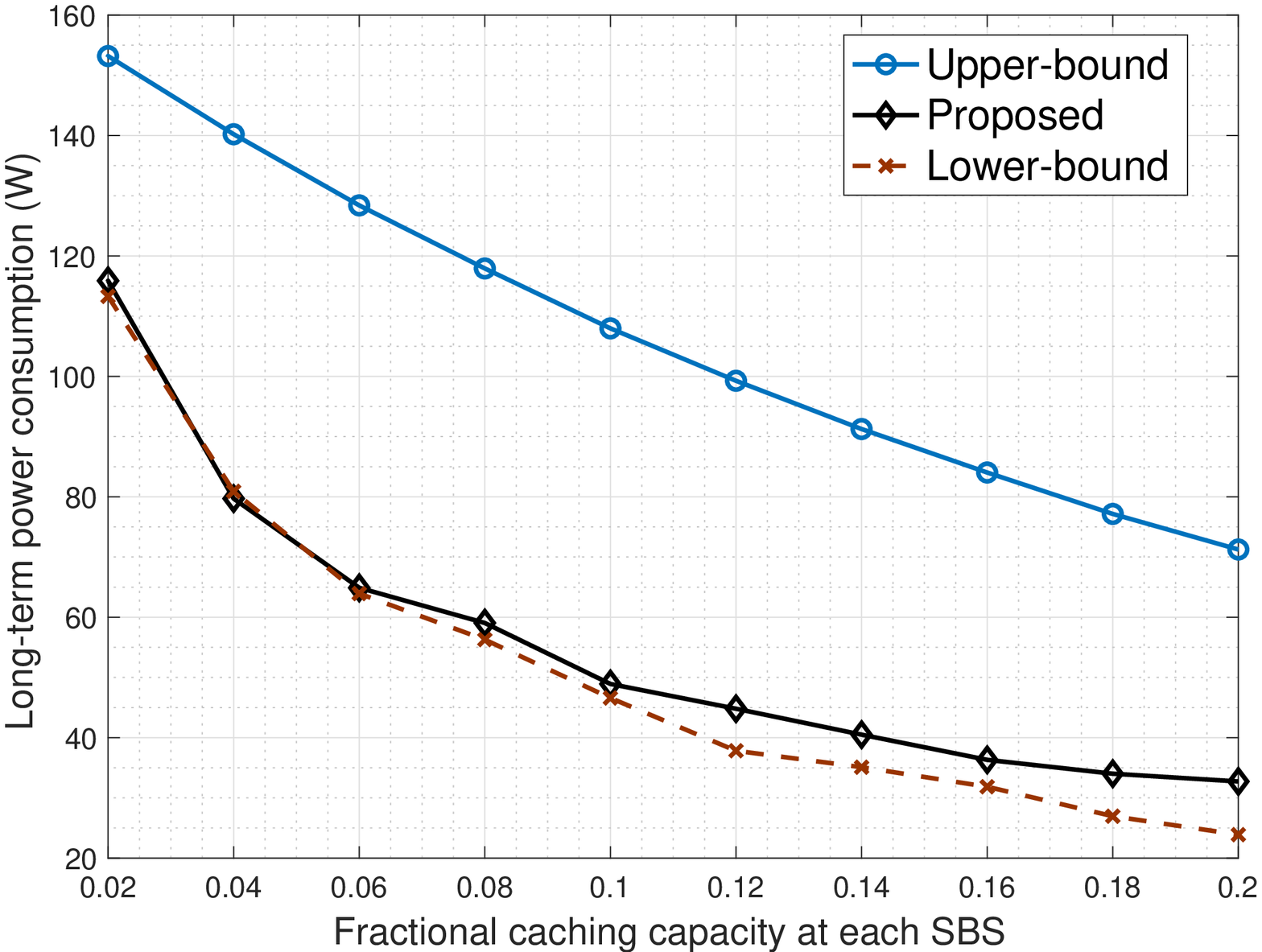}\\
  ~~~~~~~~~~~~~~~(a)	
  \end{minipage}
  \begin{minipage}{0.5\linewidth}
   \centering
  \includegraphics[scale=0.39]{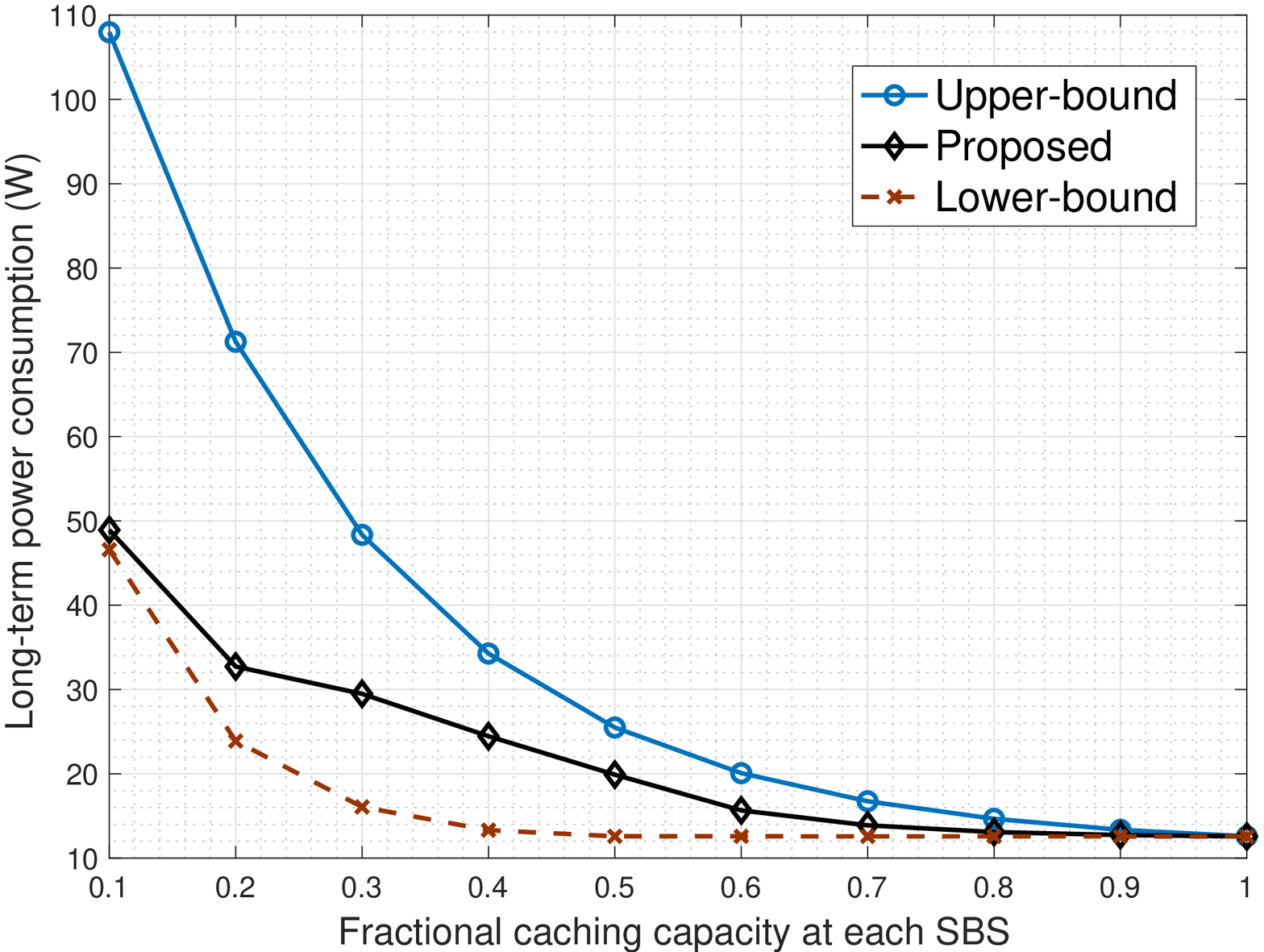} \\
  ~~~~~~~~~~~~~~~(b)	
  \end{minipage}
  \caption{The effects of fractional caching capacity.}
  \label{Fig:FCC}
\end{figure}
\subsection{Effects of Fractional Caching Capacity in SBSs}

The effects of fractional caching capacity in each SBS on the proposed framework are plotted in Fig. \ref{Fig:FCC}.
It is important to investigate caching performance under a low caching region in practical implementation because the storage size at SBS is usually somewhat restricted compared to the massive amounts of contents in the Cloud. As shown in Fig \ref{Fig:FCC}(a), when the caching capacity is within $6\%$, the proposed scheme almost matches the lower-bound scheme; when more storage is available at SBSs, the gap between the proposed scheme and the lower-bound gradually becomes larger. In general, under the low caching region, which is our main concern in practice, the proposed scheme performs very close to the lower-bound and consumes $44.3\%$ less power than the upper-bound scheme. Furthermore, the curves of the proposed scheme and baselines over the entire caching region are plotted in Fig. \ref{Fig:FCC}(b). Specifically, when fractional caching storage becomes large, the curve of the  proposed scheme gradually decreases and approaches the lower-bound. Moreover, when the fractional caching capacity is 40$\%$ or more, the power consumption of the lower-bound scheme generally remains constant. This is because the SBSs have sufficient storage to cache all  requests given the genie information. The curve of the proposed scheme merges the lower-bound in the high caching region [0.7,1] as anticipated. Indeed, a larger caching storage makes it more likely to find the requested contents in local SBSs.


\section{Conclusions}
In this paper, we have developed an efficient content caching framework by jointly considering long-term cache updating and short-term content delivery. With the goals of reducing long-term power consumption and catering to the requirements of users' QoS, a mixed timescale problem has been formulated to optimize MDS coded cache updating, multicast beamformers, and SBS clustering. Without pre-specifying any prior information about content popularity, the caching contents are allowed to be updated by making use of historical users' content requests and CSI. From the perspective of engineering implementation, we have proposed two efficient algorithms to address long-term cache updating design and short-term content delivery design, respectively. 
Furthermore, a low-complexity algorithm has been proposed by using a learning-based method to capture local content preferences. Finally, simulation results have been presented to demonstrate that the proposed caching framework outperforms extant caching and transmission schemes, and performs close to the genie-aided lower-bound under broad fronthaul bandwidths and heterogeneous content preference scenarios. 
\appendix
\subsection{\textit{Proof of Proposition 1}}   
\label{appen:A}
According to the principle of MDS decoding, the multicast traffic load for each segment of content $f$ is $\max_{b \in B_{f,t}} (1 - l_{f,b}) s_0$ \cite{liao2017coding}. Consider that each SBS always has sufficient buffering capacity for the decoding process. Our main concern is to allow the contents to be delivered continuously during content delivery, regardless of initial buffering delays.

Let $t_f$ be the buffering time of each segment during content delivery, which can be given by
\begin{align}
   t_f =\left[\frac{\max_{b \in B_{f,t}} (1 - l_{f,b}) s_0}{R_{f,t}^{\text{FH}}}-  \frac{ s_0} {R_f}\right]^+, 
\end{align}
where operator $ (\cdot)^+ = \max \{\cdot, 0\}$.
Thus, when the buffering time $t_f = 0$, we have
\begin{align}
R_{f,t}^{\text{FH}} \geq \max_{b \in \B} ~ (1 - l_{f,b})e_{f,b,t}R_f.
\end{align}
This implies that the transmission of the current segment can directly occur after users have received the previous one, i.e., no buffering time is observed during content downloading. Otherwise, the SBSs have to wait $t_f$ seconds in order to complete fetching the missing part of the current segment.
The above analysis completes the proof.

\subsection{\textit{Proof of Proposition 2}}
\label{appen:B}
{As aforementioned, constraints \eqref{h}-\eqref{l} and \eqref{eq:FH2} are equivalently reformulated. Hence, we focus on constraint \eqref{Prop2}.}
To begin, \eqref{eq:mulrat} can be reformulated as
\begin{align}
  \!R_{f,t}^{\text{FH}} \leq B_2 \log_2 (1 + \|\mb H_{b,t}^H \mb w_{f,t}\|^2/z_b^2), \forall f \in \F, b \in \mc B_{f,t}. \label{Pro2:1}
\end{align}
Clearly, for problem $\mc P_t$, the inequality in constraint \eqref{Pro2:1} should be active for the SBS $b' \in \mc B_{f,t}$ with the lowest fronthaul capacity. Otherwise, one can always find other feasible beamformers $\mb w_{f,t}$ that give rise to a smaller objective value. Moreover, constraint \eqref{Prop2} can be rewritten as
\begin{align}
    \!R_{f,t}^{\text{FH}} &\leq B_2 \log_2 (1 + \|\mb H_{b,t}^H \mb w_{f,t}\|^2/z_b^2) \notag \\&+ \tau_0(1-e_{f,b,t}), \forall f \in \F, b \in \mc B. \label{Pro2:2}
\end{align}
We consider the following two cases. Firstly, when $e_{f',b',t} = 1$, we have $b' \in \mc B_{f',t}$. In this case, for any $(f',b')$, the associated inequality in constraint \eqref{Pro2:2} is equivalent to the corresponding inequality in constraint \eqref{Pro2:1}. For the other case where $e_{f',b',t} = 0$, we have $b' \notin \mc B_{f',t}$; thus, none of the associated inequality exists in constraint \eqref{Pro2:1}; by recalling \eqref{eq:FH}, when an optimal solution to $\mc P_t$ is met, it gives rise to
$ B_2 \log_2 (1 + \|\mb H_{b',t}^H \mb w_{f',t}\|^2/z_{b'}^2) + \tau_0(1-e_{f',b',t}) \geq B_2 \log_2 (1 + \|\mb H_{b',t}^H \mb w_{f',t}\|^2/z_{b'}^2) + R_{f'} \geq R_{f',t}^{\text{FH}} $. 
As can be observed, the associated inequality in \eqref{Pro2:2} is redundant if $e_{f',b',t} = 0$ and $b' \notin \mc B_{f',t}$. Hence, $\mc P_{1,t}$ attains the same optimal value as that of $\mc P_t$. Based on the above analysis, Proposition 2 holds.

\bibliographystyle{IEEEtran}
\bibliography{references}
\begin{IEEEbiography}[{\includegraphics[width=1in,height=1.25in,clip,keepaspectratio]{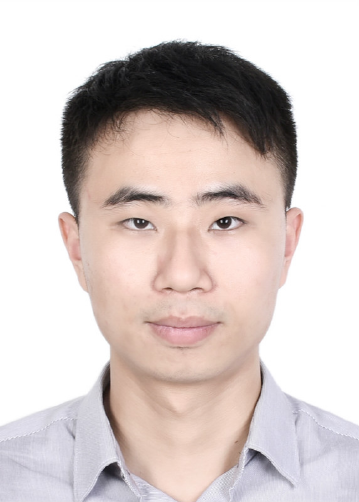}}]{Xiongwei Wu}(S'18)
received the B.Eng. in Electronic Information Engineering from University of Electronic Science and Technology of China (UESTC), Chengdu, China, in 2016. He is currently working toward the Ph.D. degree with the Chinese University of Hong Kong (CUHK), Shatin, Hong Kong SAR, China. From August 2018 to December 2018, he was a Visiting International Research Student with the University of British Columbia (UBC), Vancouver, BC, Canada. He is currently a Visiting Student Research Collaborator with Princeton University, Princeton, NJ, USA. His research interests include signal processing and resource allocation in wireless networks, decentralized optimization, and machine learning.
\end{IEEEbiography}
\begin{IEEEbiography}[{\includegraphics[width=1in,height=1.25in,clip,keepaspectratio]{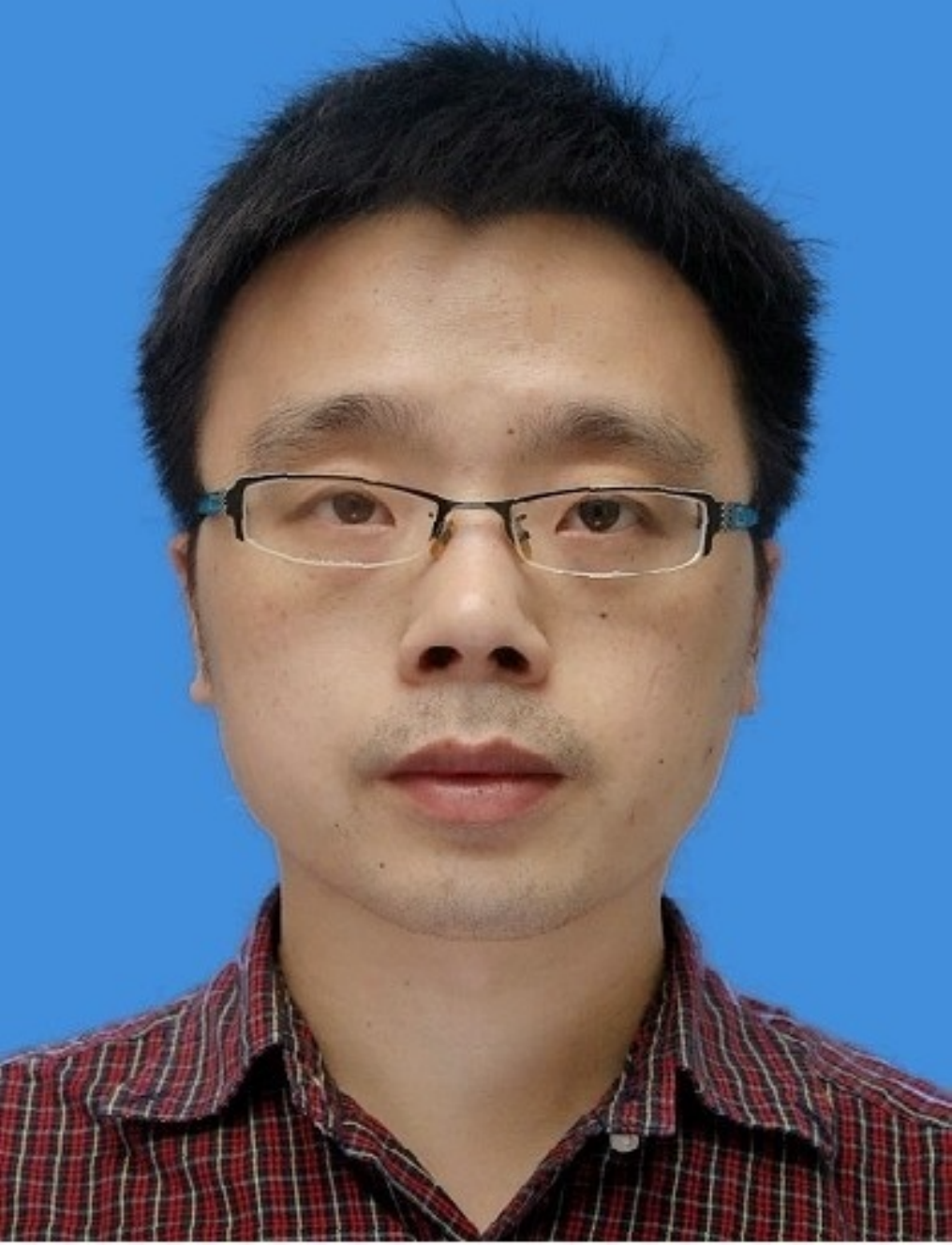}}]{Qiang Li} (M'13)
received the B.Eng. and M.Phil. degrees in Communication and Information Engineering from University of Electronic Science and Technology of China (UESTC), Chengdu, China, and the Ph.D. degree in Electronic Engineering from the Chinese University of Hong Kong (CUHK), Hong Kong, in 2005, 2008, and 2012, respectively. From August 2011 to January 2012, he was a Visiting Scholar with the University of Minnesota, Minneapolis, MN, USA. From February 2012 to October 2013, he was a Research Associate with the Department of Electronic Engineering and the Department of Systems Engineering and Engineering Management, CUHK. Since November 2013, he has been with the School of Information and Communication Engineering, UESTC, where he is currently an Associate Professor. Dr. Li is also affiliated with Peng Cheng Laboratory. His research interests include efficient optimization algorithm design for wireless communications and machine learning.

He received the First Prize Paper Award in the IEEE Signal Processing Society Postgraduate Forum Hong Kong Chapter in 2010, a Best Paper Award of IEEE PIMRC 2016, and the Best Paper Award of the IEEE Signal Processing Letters 2016.
\end{IEEEbiography}

\begin{IEEEbiography}[{\includegraphics[width=1in,height=1.25in,clip,keepaspectratio]{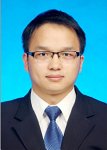}}]{Xiuhua Li} received the B.S. degree from the Honors School, Harbin Institute of Technology, Harbin, China, in 2011, the M.S. degree from the School of Electronics and Information Engineering, Harbin Institute of Technology, in 2013, and the Ph.D. degree from the Department of Electrical and Computer Engineering, The University of British Columbia, Vancouver, BC, Canada, in 2018. He joined Chongqing University through One-Hundred Talents Plan of Chongqing University in 2019. He is currently a tenure-track Assistant Professor with the School of Big Data \& Software Engineering, and the Dean of the Institute of Intelligent Network and Edge Computing associated with Key Laboratory of Dependable Service Computing in Cyber Physical Society, Chongqing University, Chongqing, China. His current research interests are 5G/B5G mobile Internet, mobile edge computing and caching, big data analytics and machine learning.
\end{IEEEbiography}

\begin{IEEEbiography}[{\includegraphics[width=1in,height=1.25in,clip,keepaspectratio]{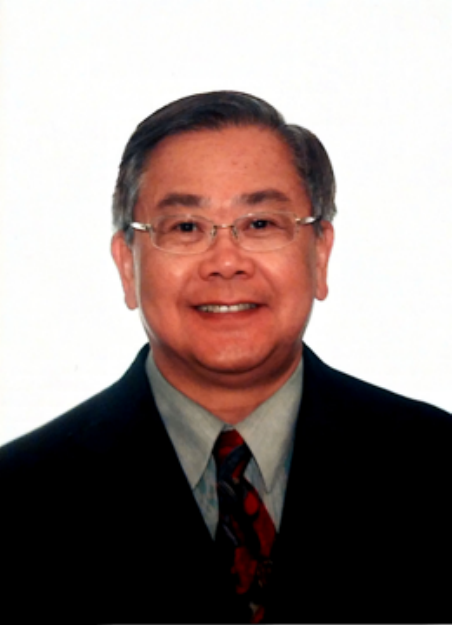}}]{Victor C. M. Leung} (S'75-M'89-SM'97-F'03) is a Distinguished Professor of Computer Science and Software Engineering at Shenzhen University. He was a Professor of Electrical and Computer Engineering and holder of the TELUS Mobility Research Chair at the University of British Columbia (UBC) when he retired from UBC in 2018 and became a Professor Emeritus.  His research is in the broad areas of wireless networks and mobile systems. He has co-authored more than 1300 journal/conference papers and book chapters. Dr. Leung is serving on the editorial boards of the IEEE Transactions on Green Communications and Networking, IEEE Transactions on Cloud Computing, IEEE Access, IEEE Network, and several other journals. He received the IEEE Vancouver Section Centennial Award, 2011 UBC Killam Research Prize, 2017 Canadian Award for Telecommunications Research, and 2018 IEEE TCGCC Distinguished Technical Achievement Recognition Award. He co-authored papers that won the 2017 IEEE ComSoc Fred W. Ellersick Prize, 2017 IEEE Systems Journal Best Paper Award, 2018 IEEE CSIM Best Journal Paper Award, and 2019 IEEE TCGCC Best Journal Paper Award.  He is a Fellow of IEEE, the Royal Society of Canada, Canadian Academy of Engineering, and Engineering Institute of Canada. He is named in the current Clarivate Analytics list of ``Highly Cited Researchers''.
\end{IEEEbiography}
\begin{IEEEbiography}[{\includegraphics[width=1in,height=1.25in,clip,keepaspectratio]{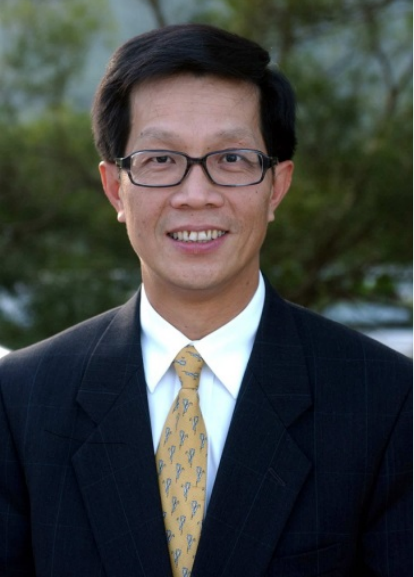}}]{P. C. Ching}(F'11)
received the B. Eng. (1st Class Honors) and Ph.D. degrees from the University of Liverpool, UK, in 1977 and 1981 respectively. From 1981 to 1982, he was Research Officer at the University of Bath, UK. In 1982, Prof. Ching returned to Hong Kong and joined the then Hong Kong Polytechnic as a lecturer. Since 1984, he has been with the Department of Electronic Engineering of the Chinese University of Hong Kong (CUHK), where he is currently Choh-Ming Li Professor of Electronic Engineering. He was Department Chairman from 1995 to 1997, Dean of Engineering from 1997 to 2003 and Head of Shaw College from 2004 to 2008. He became Director of the Shun Hing Institute of Advanced Engineering in 2004. From 2006 till end of 2014, Prof Ching was appointed as Pro-Vice-Chancellor/Vice-President of CUHK. Between 2013 to 2014, Prof. Ching also took up the Directorship of the CUHK Shenzhen Research Institute. Prof. Ching is very active in promoting professional activities, both in Hong Kong and overseas. He was a council member of the Institution of Electrical Engineers (IEE), past chairman of the IEEE Hong Kong Section, an associate editor of the IEEE Transactions on Signal Processing from 1997 to 2000 and IEEE Signal Processing Letters from 2001 to 2003. He was also a member of the Technical Committee of the IEEE Signal Processing Society from 1996 to 2004. He was appointed Editor-in-Chief of the HKIE Transactions between 2001 and 2004. He has been an Honorary Member of the editorial committee for Journal of Data Acquisition and Processing since 2000. Prof. Ching has been instrumental in organizing many international conferences in Hong Kong including the 1997 IEEE International Symposium on Circuits and Systems where he was the Vice-Chairman. He also served as Technical Program Co-Chair of the 2003 and 2016 IEEE International Conference on Acoustics, Speech and Signal Processing. Prof Ching was awarded the IEEE Third Millennium Award in 2000 and the HKIE Hall of Fame in 2010. In addition, Prof. Ching also plays an active role in community services.  He was awarded the Silver Bauhinia Star (SBS) and the Bronze Bauhinia Star (BBS) by the HKSAR Government in 2017 and 2010, respectively, in recognition of his long and distinguished public and community services.  He is presently Chairman of the Board of Directors of the Nano and Advanced Materials Institute Limited, Council Member of the Shaw Prize Foundation, as well as a Member of the Museum Advisory Committee (MAC) and the Chairperson of its Science Sub-committee. He is elected as President of the Hong Kong Academy of Engineering Sciences (HKAES) in 2018. 
\end{IEEEbiography}

\end{document}